\documentclass[aps,a4paper,superscriptaddress,twocolumn]{revtex4}
\usepackage{tensor}
\usepackage{graphicx}
\usepackage{amsmath}
\usepackage{amssymb}
\usepackage{enumerate}
\usepackage{subfigure}
\usepackage{tabularx}
\usepackage[colorlinks=true, pdfstartview=FitV, linkcolor=blue, citecolor=red, urlcolor=black, breaklinks=true]{hyperref}
\newcommand{\be}{\begin{equation}}
\newcommand{\ee}{\end{equation}}
\newcommand{\ben}{\begin{eqnarray}}
\newcommand{\een}{\end{eqnarray}}
\newcommand{\bes}{\begin{subequations}}
\newcommand{\ees}{\end{subequations}}
\def\bal#1\eal{\begin{align}#1\end{align}}

\newcommand{\LL}{{\mathcal L}}

\newcommand{\A}{\mathcal{A}}
\newcommand{\D}{\mathcal{D}}
\newcommand{\F}{\mathcal{F}}
\newcommand{\Hc}{\mathcal{H}}
\newcommand{\K}{\mathcal{K}}
\newcommand{\M}{\mathcal{M}}
\newcommand{\Pc}{\mathcal{P}}
\newcommand{\veps}{\varepsilon}
\begin{document}
\title{Bimagnetic monopoles}

\author{D. Bazeia}
\affiliation{Departamento de F\'\i sica, Universidade Federal da Para\'\i ba, 58051-970 Jo\~ao Pessoa, PB, Brazil}
\author{M.A. Marques}
\affiliation{Departamento de F\'\i sica, Universidade Federal da Para\'\i ba, 58051-970 Jo\~ao Pessoa, PB, Brazil}
\author{R. Menezes}
\affiliation{Departamento de F\'\i sica, Universidade Federal da Para\'\i ba, 58051-970 Jo\~ao Pessoa, PB, Brazil}
\affiliation{Departamento de Ci\^encias Exatas, Universidade Federal da Para\'\i ba, 58297-000 Rio Tinto, PB, Brazil}

\begin{abstract}{\smallskip}
Motivated by recent results on small and hollow magnetic monopoles and on core and shell bimagnetic nanoparticles, we propose the construction of bimagnetic monopoles, which are structures that accommodate a magnetic monopole inside another magnetic monopole.\\  
\end{abstract}

\maketitle

\section{Introduction}

In 1974, two independent pioneer studies by 't Hooft \cite{thooft} and Polyakov \cite{polyakov} unveiled the presence of magnetic monopoles in a relativistic model described by a set of gauge and scalar fields controlled by the non Abelian $SU(2)$ symmetry. Soon after, two other investigations described an interesting way to reduce the corresponding equations of motion to first order differential equations that engender minimum energy monopole solutions \cite{ps,bogo}. The method is usually referred to as the BPS procedure, to honor the authors of Refs.~\cite{ps,bogo}.

The study of magnetic monopoles of the type described above is hard, because it requires the presence of several degrees of freedom, controlled by the gauge and scalar fields that interact under a local and non Abelian symmetry in three spatial dimensions; see, e.g., \cite{V,MS,Sh} for revisions on the subject of magnetic monopoles. In this sense, the presence of the BPS procedure is welcome since it leads to first order differential equations that support stable minimum energy solutions that are somehow simpler to be obtained.

Other studies on magnetic monopoles have been motivated by the addition of internal structure to the magnetic configuration. An interesting way to do this is with the enhancement of the symmetry, as in the investigation implemented before in \cite{W,Shi}, in which the symmetry is modified to accommodate new degrees of freedom, to show how a conventional topological defect may acquire other features. Another possibility appeared recently in \cite{HS}, in which the authors study the presence of colormagnetic structures in dense quark matter that can be used to model the interior of compact stars. Other recent works on magnetic monopoles appeared before in Refs.~\cite{M1,M2,M3,M4,M5,Ba1,Ba2,Ba3}. In particular, in \cite{Ba2} we changed the $SU(2)$ symmetry to the case of $SU(2)\times Z_2$, with the inclusion of an extra neutral scalar field to investigate the construction of magnetic monopoles with internal structure, in \cite{M5} the authors studied the existence of fermionic zero modes in the background of a Kaluza-Klein monopole and also, in \cite{Ba3} we discovered new possibilities, in which the magnetic structures may have small and hollow features. The presence of these novel monopoles \cite{Ba3} motivated us to study the possibility that we describe in the current work, concerning the construction of bimagnetic monopoles, that are structures composed of a magnetic monopole which is nested inside another magnetic monopole.

The bimagnetic structure is of current interest in high energy physics and may find applications in other areas of nonlinear science, in particular in condensed matter. Evidently, it gains further importance if one reminds the presence of magnetic monopoles in a class of exotic magnetic materials, which are known collectively as spin ice \cite{SI1,SI2,SI3,SI4}. In this case, the dipole moments of the localized structures may fractionalize in the form of monopoles, and more, besides having magnetic charge, they may also be endowed with an electric dipole \cite{SI5}. In this sense, in spin ice the magnetic configurations may engender a richer internal structure. But there are more, we can also construct bimagnetic materials, in which a magnetic component, usually called core, may be covered by another magnetic component, called shell. This has been studied recently in Refs.~\cite{Bi1,Bi2}, and these composite structures have inspired us to investigate the possibility to construct other configurations, in particular the case in which a magnetic monopole can be nested inside another magnetic monopole. 

The magnetic structures that we construct here are spatial configurations that engender spherical symmetry, and this is not  present in the structures that appeared before in Refs.~\cite{Bi1,Bi2}. In this sense, we are still further from the bimagnetic objects that appear in condensed matter, but we think that the current investigation offers an interesting theoretical construction of magnetic monopoles with internal structure. In particular, we will also explore the possibility of constructing a shell on top of another shell, with an empty core, and this is certainly simpler to realize experimentally. Furthermore, since the bimagnetic structures to be investigated here require the doubling of the degrees of freedom, changing the standard $SU(2)$ symmetry originally considered in Refs.~\cite{thooft,polyakov} to the case of $SU(2)\times SU(2)$, one may naturally ask whether these bimagnetic structures play a hole in models that deal with issues beyond the Standard Model, with semisimple groups like the Pati-Salam model proposed in Ref.~\cite{PS}, for instance. Models of this type are based on the left-right symmetry \cite{PS,LR1,LR2}, which operates under the presence of the $SU(2)\times SU(2)$ subgroup, and this is the symmetry we need to construct a magnetic monopole inside another magnetic monopole.

The enhancement of symmetries to study the entrapment of topological structures can be implemented in many different contexts, in particular in the simpler case involving discrete and/or Abelian gauge symmetries. Investigations focusing on this issue appeared before, for instance, in \cite{D1,D2}, and also in \cite{D3}, where a model with $Z_2\times Z_3$ symmetry was considered to describe the entrapment of a hexagonal network of topological defects inside a domain wall. Models involving the global and local Abelian $U(1)$ symmetry were also investigated in \cite{Sut,Shif,Ba4} and in references therein. The investigations \cite{D1,D2,D3,Sut,Shif,Ba4} can be added to the set of works dealing with non Abelian symmetries \cite{Shi,HS,M1,Ba2,Ba3} to motivate other studies on the construction of topological solutions with internal structure.

To comply with the construction of bimagnetic monopoles, in Sec.~\ref{sec:mod} we start describing the system and developing the BPS procedure on general grounds. We then investigate specific examples in Sec.~\ref{sec:exe}, considering the entrapment of a standard monopole inside a hollow monopole and also, describing the case where a small monopole can be nested inside a hollow monopole. In order to approach the practical realization of the results of the current work, we also constructed a hollow monopole inside another hollow monopole. Since the hollow monopole has an empty core, we are then dealing with a shell on top of another shell, with an empty core. In this way, we are still keeping spherical symmetry, but now, concerning the practical realization of the corresponding composite structure, we are circumventing the presence of singularity at the center of the solution, approaching a more feasible possibility. We end the work in Sec.~\ref{sec:end}, with the inclusion of comments and perspectives of future works.

%%%%%%%%%%%%%%%%%%%%
\section{The Model}
\label{sec:mod}

The model is defined in $(3,1)$ spacetime dimensions and we consider the Lagrangian density
\be\label{lmodel}
\begin{aligned}
\LL &= - \frac{P(|\chi|)}{4}F^{a}_{\mu\nu}F^{a\mu\nu} -\frac{M(|\chi|)}{2} D_\mu \phi^a D^\mu \phi^a \\
&\hspace{4mm} - \frac{\Pc(|\chi|)}{4}\F^{a}_{\mu\nu}\F^{a\mu\nu} -\frac{\M(|\chi|)}{2} \D_\mu \chi^a \D^\mu \chi^a\\
&\hspace{4mm} - V(|\phi|,|\chi|).
\end{aligned}
\ee
It engenders $SU(2)\times SU(2)$ symmetry. Here, $\phi^a$ represents a triplet of real scalar fields that is coupled to the gauge field $A^a_\mu$ under the $SU(2)$ symmetry. The other $SU(2)$ symmetry describes the fields $\chi^a$ and $\A^a_\mu$. The covariant derivatives are denoted by $D_\mu \phi^a = \partial_\mu\phi^a + g\veps^{abc}A^b_\mu\phi^c$ and $\D_\mu \chi^a = \partial_\mu\chi^a + q\veps^{abc}\A^b_\mu\chi^c$, and the field strength tensors are $F^a_{\mu\nu} = \partial_\mu A^a_\nu - \partial_\nu A^a_\mu + g\veps^{abc}A^b_\mu A^c_\nu$ and $\F^a_{\mu\nu} = \partial_\mu \A^a_\nu - \partial_\nu \A^a_\mu + q\veps^{abc}\A^b_\mu \A^c_\nu$. In our model, $P(|\chi|)$ and $\Pc(|\chi|)$ represents the magnetic permeabilities, and the functions $M(|\chi|)$ and $\M(|\chi|)$ control the kinetic terms of the scalar fields, and we suppose they are all nonnegative functions of $|\chi|$. Also, $g$ and $q$ are the coupling constants, the indices $a,b,c=1,2,3$ are used to represent the $SU(2)$ symmetries and the greek letters $\mu,\nu=0,1,2,3$ stand for the spacetime indices. We use the Minkowski metric, $\eta_{\mu\nu} = \textrm{diag}(-,+,+,+)$, with natural units, $\hbar=c=1$.

The field equations associated to the Lagrangian density in Eq.~\eqref{lmodel} are given by
\bes\label{geom}
\begin{align}
 D_\mu\left(M D^\mu \phi^a\right) &=V_{\phi^a},\\
  D_\mu\left(\M D^\mu \chi^a\right) &=\frac{P_{\chi^a}}{4}F^{b}_{\mu\nu}F^{b\mu\nu} + \frac{M_{\chi^a}}{2} D_\mu \phi^b D^\mu \phi^b \nonumber\\
                                    &\hspace{4mm}\frac{\Pc_{\chi^a}}{4}\F^{b}_{\mu\nu}\F^{b\mu\nu} + \frac{\M_{\chi^a}}{2} \D_\mu \chi^b \D^\mu \chi^b +  V_{\chi^a},\\ \label{meqsc}
 D_\mu\left(PF^{a\mu\nu}\right) &=  gM\,\veps^{abc}\phi^b D^\nu \phi^c,\\
 \D_\mu\left(\Pc\F^{a\mu\nu}\right) &=  q\M\,\veps^{abc}\chi^b \D^\nu \chi^c,
\end{align}
\ees
in which we have $D_\mu F^{a\mu\nu} = \partial_\mu F^{a\mu\nu} + g\veps^{abc}A^b_\mu F^{c\mu\nu}$, $\D_\mu \F^{a\mu\nu} = \partial_\mu \F^{a\mu\nu} + q\veps^{abc}\A^b_\mu \F^{c\mu\nu}$, $V_{\phi^a} = \partial V/\partial\phi^a$ and $V_{\chi^a} = \partial V/\partial\chi^a$, etc.

To investigate the presence of monopoles, we take static configurations with $A_0=\A_0=0$, which yields to null electric fields, and the hedgehog ansatz
\bes\label{ansatz}
\bal
\phi^a &= \frac{x_a}{r} H(r) \quad \text{and} \quad A_i^a = \veps_{aib}\frac{x_b}{gr^2}(1-K(r)),\\
\chi^a &= \frac{x_a}{r} \Hc(r) \quad \text{and} \quad \A_i^a = \veps_{aib}\frac{x_b}{qr^2}(1-\K(r)),
\eal
\ees
with the boundary conditions
\be\label{bcond}
\begin{aligned}
H(0)&=0, & K(0)&=1,\\
\Hc(0)&=0, & \K(0)&=1,\\
H(\infty) &\to \pm v, & K(\infty) &\to 0,\\
\Hc(\infty) &\to \pm w, & \K(\infty) &\to 0,
\end{aligned}
\ee
where $v$ and $w$ are positive real numbers. As usual, the boundary conditions for $H, K, \Hc,$ and $\K$ at the origin have to be used to leave no room for singular behavior for the corresponding fields. In this case, the presence of spherical symmetry changes the equations of motion \eqref{geom} to 
\bes\label{geomansatz}
\begin{align}
\frac{1}{r^2}\left(r^2M H^\prime\right)^\prime &= \frac{2M H K^2}{r^2} + V_{H},   \\
\frac{1}{r^2}\left(r^2\M \Hc^\prime\right)^\prime &= \frac{2\M \Hc \K^2}{r^2} +\frac{P_\Hc}{2} \left(\frac{2{K^\prime}^2}{g^2r^2} + \frac{(1-K^2)^2}{g^2r^4}\right)\nonumber\\
                                  &\hspace{4mm} +\frac{M_\Hc}{2}\left({H^\prime}^2 + \frac{2H^2K^2}{r^2}\right)  \nonumber \\
                                 &\hspace{4mm} +\frac{\Pc_\Hc}{2} \left(\frac{2{\K^\prime}^2}{q^2r^2} + \frac{(1-\K^2)^2}{q^2r^4}\right)\nonumber\\
                                  &\hspace{4mm} +\frac{\M_\Hc}{2}\left({\Hc^\prime}^2 + \frac{2\Hc^2\K^2}{r^2}\right) + V_{\Hc},\\
r^2\left(P K^\prime\right)^\prime &= K\left(M g^2r^2H^2 -P\,(1-K^2)\right),\\
r^2\left(\Pc \K^\prime\right)^\prime &= \K\left(\M q^2r^2\Hc^2 -\Pc\,(1-\K^2)\right),
\end{align}
\ees
where the prime stands for the derivative with respect to radial coordinate $r$. The above equations are second order differential equations that couple to each other via the several functions involved in the problem, so they are hard to solve. For this reason, it is convenient to use the BPS procedure to find the first order differential equations that solve the equations of motion. To do so, we write the energy density of the static field configurations in the form
\be\label{rho}
\begin{aligned}
\rho &= \frac{P(|\Hc|)}{2} \left(\frac{2{K^\prime}^2}{g^2r^2} + \frac{(1-K^2)^2}{g^2r^4}\right)\\
&\hspace{4mm} + \frac{M(|\Hc|)}{2}\left({H^\prime}^2 + \frac{2H^2K^2}{r^2}\right)\\
&\hspace{4mm} + \frac{\Pc(|\Hc|)}{2} \left(\frac{2{\K^\prime}^2}{q^2r^2} + \frac{(1-\K^2)^2}{q^2r^4}\right)\\     
&\hspace{4mm} + \frac{\M(|\Hc|)}{2}\left({\Hc^\prime}^2 + \frac{2\Hc^2\K^2}{r^2}\right) + V(|H|,|\Hc|).
\end{aligned}
\ee
The idea is to proceed as in \cite{Ba3} and use $M(|\Hc|)=1/P(|\Hc|)$ and $\M(|\Hc|) = 1/\Pc(|\Hc|)$ to write 
\be
\begin{aligned}
\rho &=\frac{P(|\Hc|)}{2} \left(\frac{H^\prime}{P(|\Hc|)} \mp \frac{1-K^2}{gr^2}\right)^2 \\
&\hspace{4mm} + P(|\Hc|)\left( \frac{K^\prime}{gr} \pm \frac{HK}{rP(|\Hc|)}\right)^2 \\
&\hspace{4mm} + \frac{\Pc(|\Hc|)}{2} \left(\frac{\Hc^\prime}{\Pc(|\Hc|)} \mp \frac{1-\K^2}{qr^2}\right)^2 \\
&\hspace{4mm} + \Pc(|\Hc|)\left( \frac{\K^\prime}{qr} \pm \frac{\Hc\K}{r\Pc(|\Hc|)}\right)^2 + V(|H|,|\Hc|)\\
&\hspace{4mm}  \pm \frac{1}{r^2}\left(\frac{\left(1-K^2\right)H}{g} + \frac{\left(1-\K^2\right)\Hc}{q}\right)^\prime.
\end{aligned}
\ee
We follow the suggestion introduced in Ref.~\cite{ps} and take the potential as $V(|\phi|,|\chi|)=0$. In this case, since the first four terms in the above energy density are nonnegative, we have that the energy is bounded, i.e., $E\geq E_B$, where
\be\label{ebogo}
E_B = \frac{4\pi v}{g} + \frac{4\pi w}{q}.
\ee
If the solutions satisfy the first order equations
\bes\label{foh}
\bal
\Hc^\prime &=\pm \frac{\Pc(|\Hc|)(1-\K^2)}{qr^2},\\
\K^\prime &=\mp \frac{q\Hc\K}{\Pc(|\Hc|)},
\eal
\ees
and
\bes\label{fov}
\bal
H^\prime &=\pm \frac{P(|\Hc|)(1-K^2)}{gr^2},\\
K^\prime &=\mp \frac{gHK}{P(|\Hc|)},
\eal
\ees
the energy is minimized to $E=E_B$, given by Eq.~\eqref{ebogo}, so the solutions of the first order equations are stable against decay into nontrivial lower energy configurations. We can show that solutions of the the first order equations \eqref{foh} and \eqref{fov} are also solutions of the equations of motion \eqref{geomansatz} for $V(|\phi|,|\chi|)=0$. Furthermore, both pair of equations with the upper signs are related to the lower signs ones through the change $\Hc(r)\to -\Hc(r)$ and $H(r)\to-H(r)$.

One can note that the first order equations \eqref{foh} only involve $\Hc(r)$ and $\K(r)$; they do not couple to the functions $H(r)$ and $K(r)$. Therefore, we first choose $\Pc(|\chi|)$ and solve Eqs.~\eqref{foh} to find the profile of $\Hc(r)$ and $\K(r)$, which we call the core monopole. We then substitute $\Hc(r)$ in Eqs.~\eqref{fov} and choose $P(|\chi|)$ to calculate $H(r)$ and $K(r)$, which we call the shell monopole. In this sense, the magnetic solution that appears from Eqs.~\eqref{foh} acts as a source for Eqs.~\eqref{fov}, that can form another magnetic structure. We may then have a composite of bimagnetic structure. Indeed, one can use the aforementioned first order equations in Eq.~\eqref{rho} to distinguish the energy density of the two substructures. It can be written in the form
\be
\rho = \rho_c + \rho_s,
\ee
where $\rho_c$ and $\rho_s$ represent the energy densities of the core and shell components of the bimagnetic structure. They have the forms
\be\label{rhod}
\begin{split}
\rho_c &= \frac{2\Pc(|\Hc|){\K^\prime}^2}{q^2r^2} + \frac{{\Hc^\prime}^2}{\Pc(|\Hc|)}\\
&= \frac{2\Hc^2\K^2}{r^2 \Pc(|\Hc|)} + \frac{\Pc(|\Hc|)(1-\K^2)^2}{q^2r^4}
\end{split}
\ee
and
\be\label{rhoc}
\begin{split}
\rho_s &= \frac{2P(|\Hc|){K^\prime}^2}{g^2r^2} + \frac{{H^\prime}^2}{P(|\Hc|)}\\
&= \frac{2H^2K^2}{r^2 P(|\Hc|)} + \frac{P(|\Hc|)(1-K^2)^2}{g^2r^4}.
\end{split}
\ee
It is worth highlighting here that the BPS procedure leads to a fixed energy for each component, given by Eq.~\eqref{ebogo}, despite the form of the functions $\Pc(|\Hc|)$ and $P(|\Hc|)$.

%%%%%%%%%%%%%%%%%%%%%%%%%%
\section{Examples}
\label{sec:exe}

%%%%%%%%%%%%%%%%%%%%%%%%%%%%%%
The next step is to use the above results to illustrate the general procedure with some examples. In order to prepare the model for numerical investigation, we first rescale some quantities as follows
\be
\begin{aligned}
\phi^a &\to v \phi^a, & A^a_\mu &\to v A^a_\mu, & r &\to (g v)^{-1} r,\\
\chi^a &\to v \chi^a, & \A^a_\mu &\to v \A^a_\mu, & \LL &\to g^2 v^4 \LL.
\end{aligned}
\ee
This leads us to work with dimensionless fields and radial coordinate. We also take $g=v=1$ and consider the upper signs in the first order equations \eqref{foh} and \eqref{fov}, for simplicity.

%%%%%%%%%%%%%%%%%%%%%%%%%%%%%%
\subsection{Shell on standard core monopole}
\begin{figure}[t!]
\centering
\includegraphics[width=4cm]{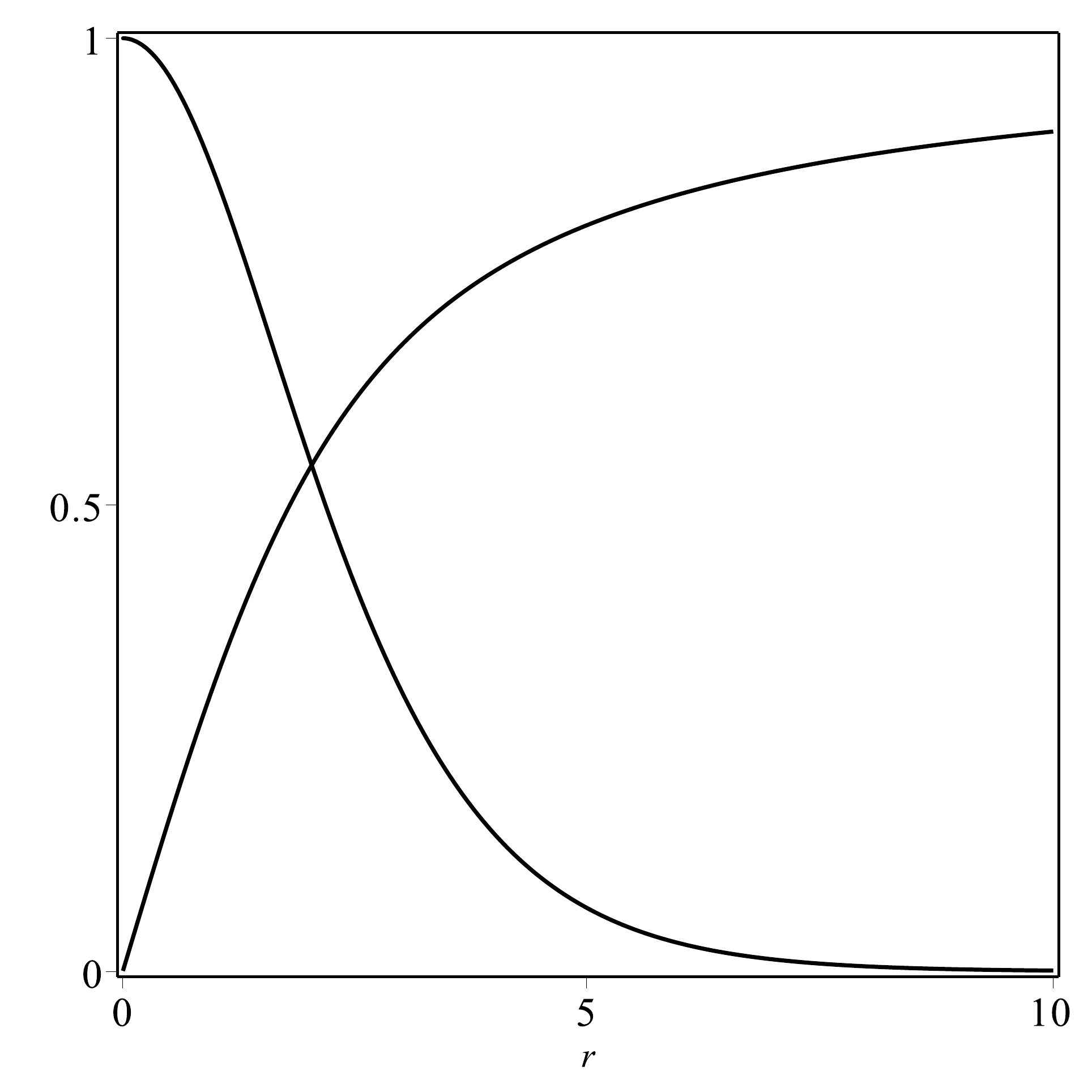}
\includegraphics[width=4cm]{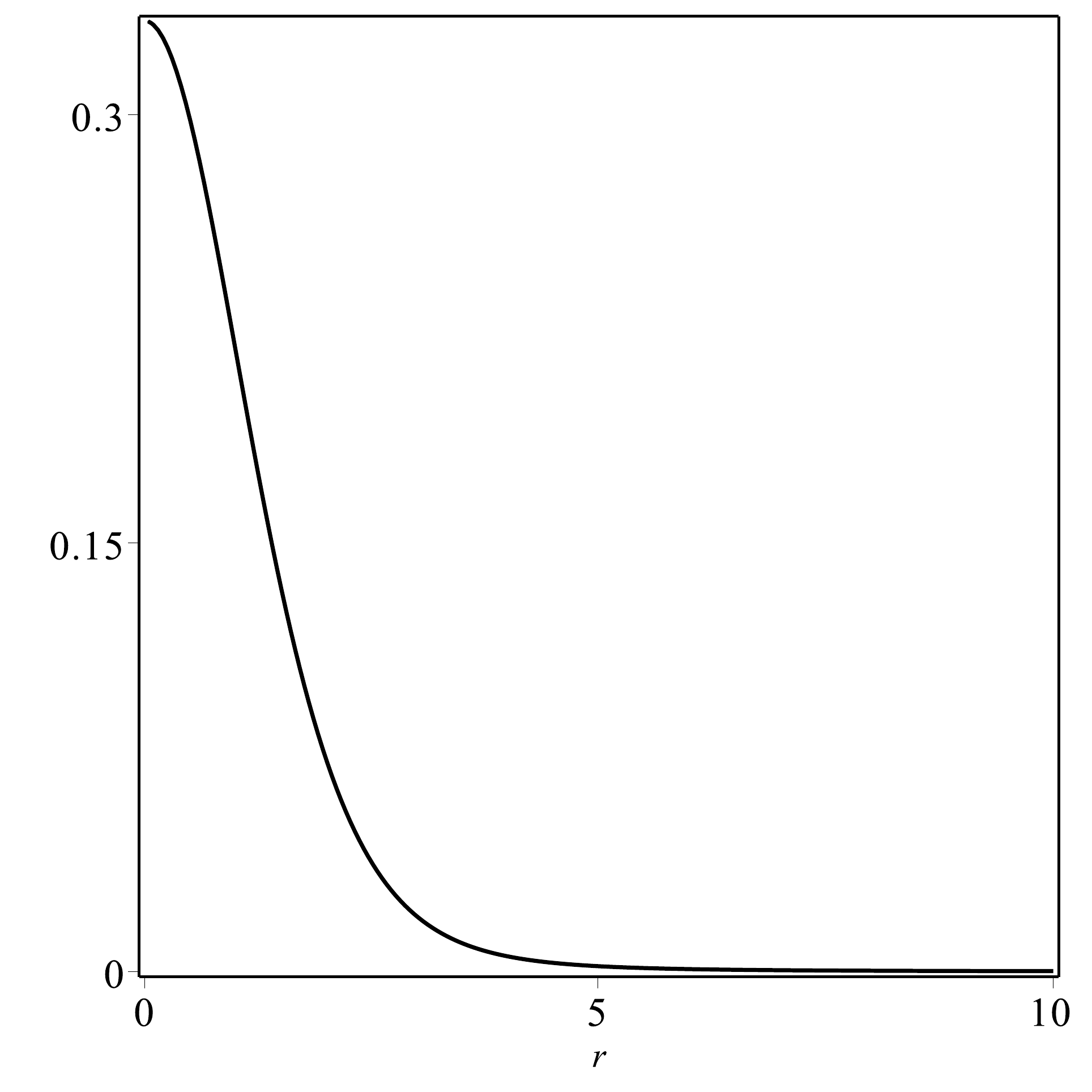}
\caption{In the left panel, we show the solutions $\K(r)$ (descending line) and $\Hc(r)$ (ascending line) that appear in Eq.~\eqref{solstd}. In the right panel we display the energy density \eqref{rhostd}.}
\label{fig1}
\end{figure}

As the first example, let us consider the core structure to be in the standard form, as proposed by 't Hooft and Polyakov \cite{thooft,polyakov}. This implies that
$\Pc(|\chi|)=1$. In this case, the first order equations \eqref{foh} become
\bes
\bal
\Hc^\prime &= \frac{(1-\K^2)}{qr^2},\\
\K^\prime &=- q\Hc\K,
\eal
\ees
They admit the analytical solutions \cite{ps}
\be\label{solstd}
\Hc(r)= \coth(r) - \frac{1}{r} \quad\text{and}\quad \K(r) = r\,\textrm{csch}(r),
\ee
where we have considered $q=w=1$, for simplicity. The energy density of the core monopole is obtained from Eq.~\eqref{rhod}, which leads to
\be\label{rhostd}
\rho_c(r) = \frac{\left(r^2\,\textrm{csch}^2(r)\!-\!1\right)^2}{r^4}\! +\! \frac{2\,\textrm{csch}^2(r)\left(r\coth(r)\!-\!1\right)^2}{r^2}.
\ee
In Fig.~\ref{fig1}, we display the solutions \eqref{solstd} and the above energy density.
%%%%%%%%%%%%%%%%%%%%%%%
%%%%%%%%%%%%%%%%%%%%%%%

To find the shell structure, we take the magnetic permeability in the form
\be\label{p1}
P(|\chi|) = {1}/{|\chi|^\alpha},
\ee
with $\alpha$ real and positive. In this case, the first order equations \eqref{fov} become
\bes\label{fov1}
\bal
H^\prime &=\frac{(1-K^2)}{(r\coth(r)-1)^\alpha r^{2-\alpha}},\\
K^\prime &=-\frac{HK(r\coth(r)-1)^\alpha}{r^\alpha},
\eal
\ees
Unfortunately, we have been unable to find analytical solutions for the above equations. However, before using numerical procedures, we can evaluate their behavior near the origin by taking $H(r)= H_o(r)$ and $K(r)=1-K_o(r)$. In this case, the most important contribution to $H_o$ and $K_o$ are: $H_o\propto r^{\left(\sqrt{\alpha^2 + 2\alpha + 9}-\alpha -1\right)/2}$ and $K_o\propto r^{\left(\sqrt{\alpha^2 + 2\alpha + 9}+\alpha +1\right)/2}$. A similar procedure can be done for the asymptotic behavior. By considering $H(r) = 1-H_{asy}(r)$ and $K(r) = K_{asy}$, we get $H_{asy}\propto r^{-1}$ and $K_{asy}(r) \propto e^{-r}$. 

%%%%%%%%%%%%%%%%%%%%%%%
\begin{figure}[t!]
\centering
\includegraphics[width=4cm]{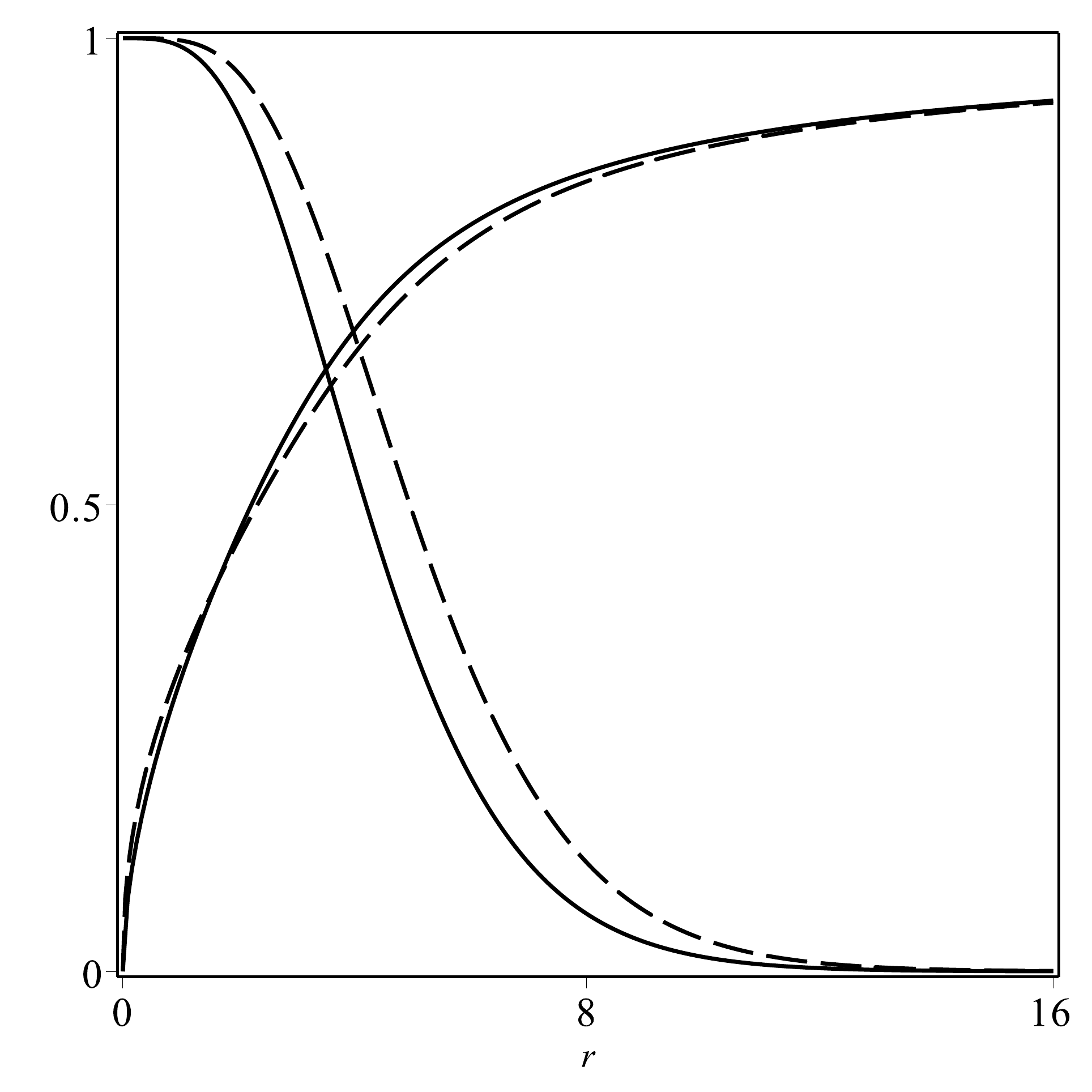}
\includegraphics[width=4cm]{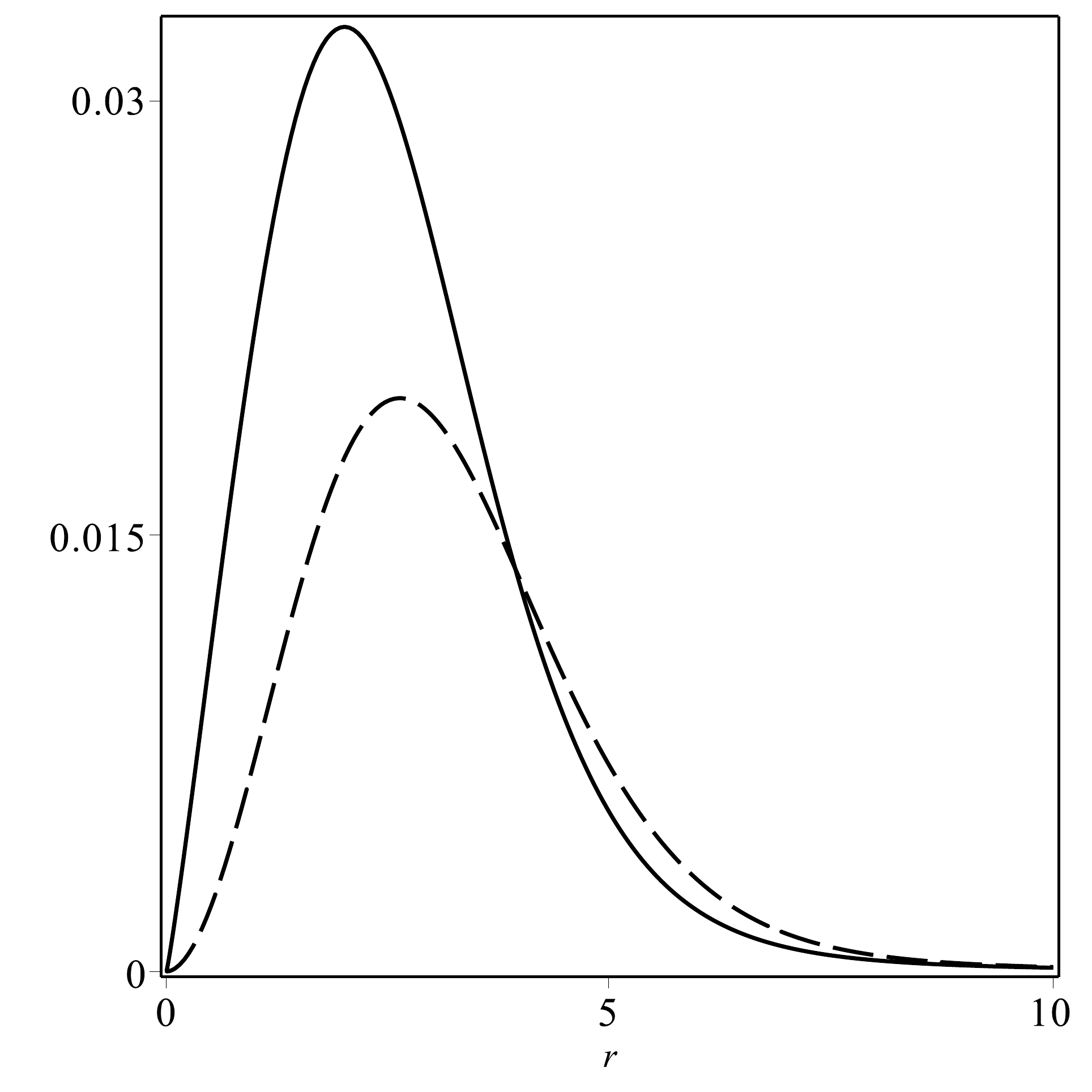}
\caption{In the left panel, we show the solutions $K(r)$ (descending line) and $H(r)$ (ascending line) of Eqs.~\eqref{fov1}, and in the right panel we display their energy density \eqref{rhoc1}. The solid lines represent the case $\alpha=2$ and the dashed ones stand for $\alpha=3$.}
\label{fig2}
\end{figure} 
%%%%%%%%%%%%%%%%%%%%%%%
%%%%%%%%%%%%%%%%%%%%%%%
\begin{figure}[t!]
\centering
\includegraphics[width=5cm]{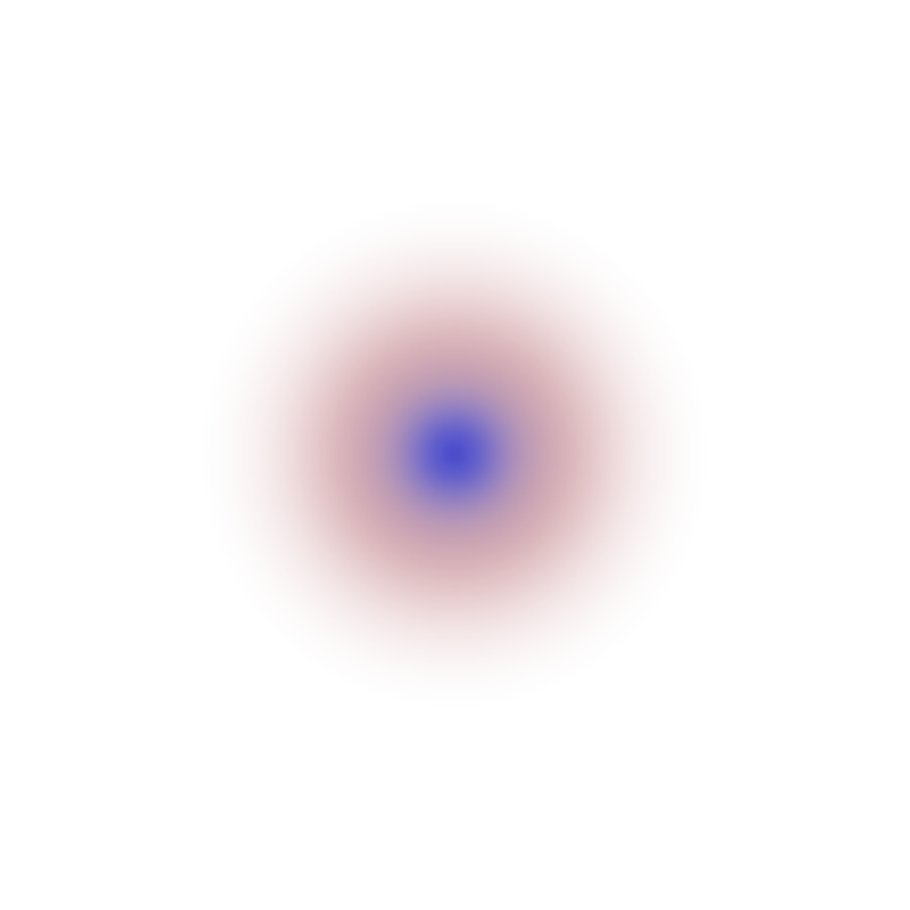}
\caption{A planar section of the energy density passing through the center of the structure.  The color blue describes the core of the structure, which is represented by the standard monopole with the energy density shown in Eq.~\eqref{rhostd}. The color red describes the shell, which is represented by the hollow monopole, with the energy density shown in Eq.~\eqref{rhoc1}, for $\alpha=3$.}
\label{fig3}
\end{figure} 
%%%%%%%%%%%%%%%%%%%%%%%

The energy density of the shell structure is given by Eq.~\eqref{rhoc} and has the form
\be\label{rhoc1}
\begin{split}
\rho_s(r) &= \frac{2{K^\prime}^2}{(r\coth(r)-1)^\alpha r^{2-\alpha}} + \frac{{H^\prime}^2(r\coth(r)-1)^\alpha}{r^\alpha}\\
&= \frac{2H^2K^2(r\coth(r)-1)^\alpha}{r^{2+\alpha}} + \frac{(1-K^2)^2}{r^{4-\alpha}(r\coth(r)-1)^\alpha}.
\end{split}
\ee
By using the behavior of the solutions around the origin, one can show that $\rho_s(r\approx0) \approx r^{\sqrt{\alpha^2 + 2\alpha + 9}-3}$. Therefore, the parameter $\alpha$ controls the way the energy density of the shell configuration behaves at its core, so we use it to appropriately describe the solution.

We use numerical methods and depict, in Fig.~\ref{fig2}, the solutions of Eqs.~\eqref{fov1} and the energy density given above, for $\alpha=2$ and $3$. The structure presents a hole around its core, which is controlled by $\alpha$ and increases as $\alpha$ also increases. This is similar to the hollow monopole found in Ref.~\cite{Ba3}. Here, however, the shell structure coexists with the core structure, whose energy density is displayed in Fig.~\ref{fig1}.

In order to highlight the result, in Fig.~\ref{fig3} we display a planar section of the energy density of the bimagnetic structure, passing through its center. We depict the energy density that appears in Eq.~\eqref{rhostd} in blue, and that shown in Eq.~\eqref{rhoc1} in red,
for $\alpha=3$. The blue core shows the standard monopole, and the red shell displays the hollow monopole. In fact, we have a single structure, which is composed of two distinct substructures and can be seen as a bimagnetic monopole.
 
%%%%%%%%%%%%%%%%%%%%%%%
\begin{figure}[htb!]
\centering
\includegraphics[width=4cm]{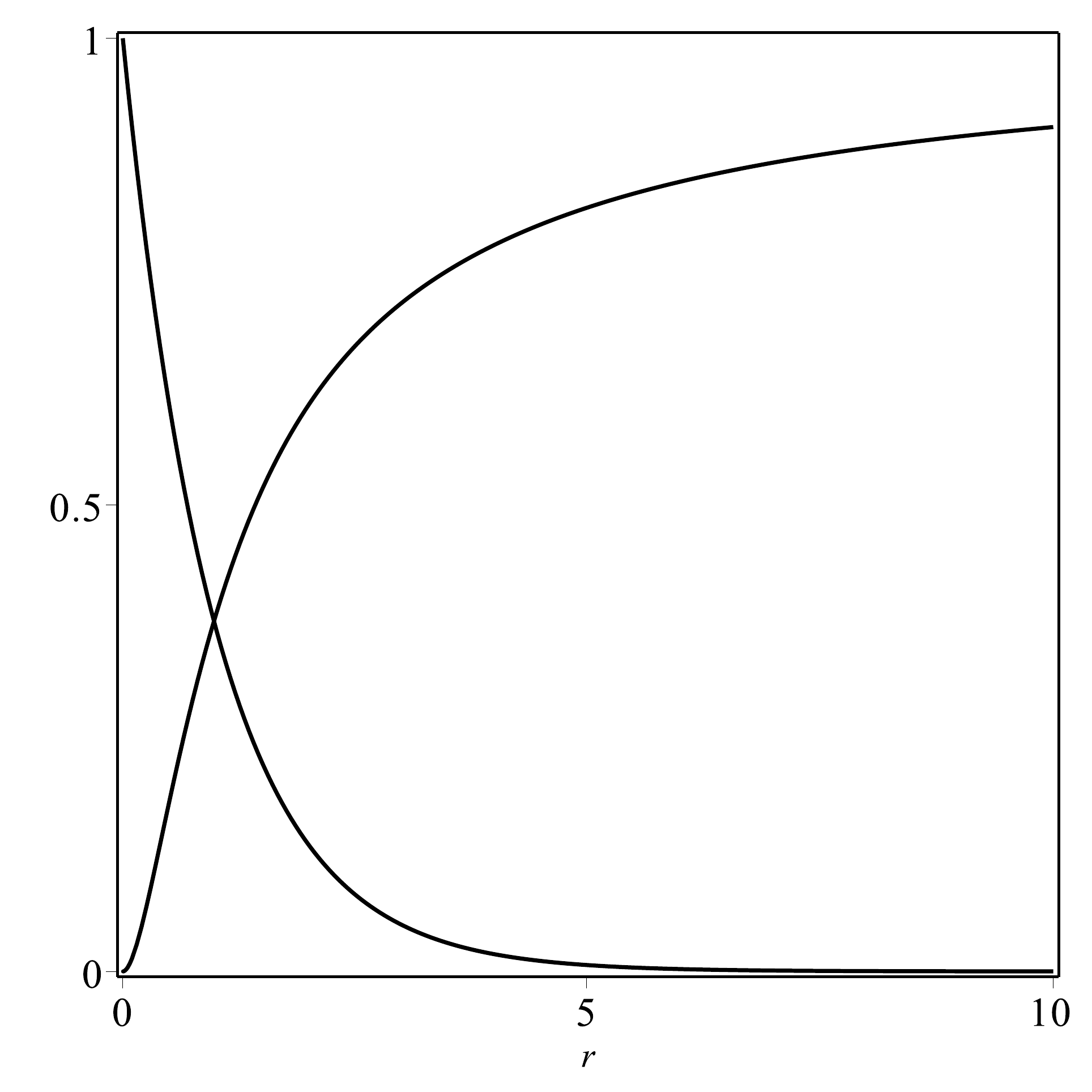}
\includegraphics[width=4cm]{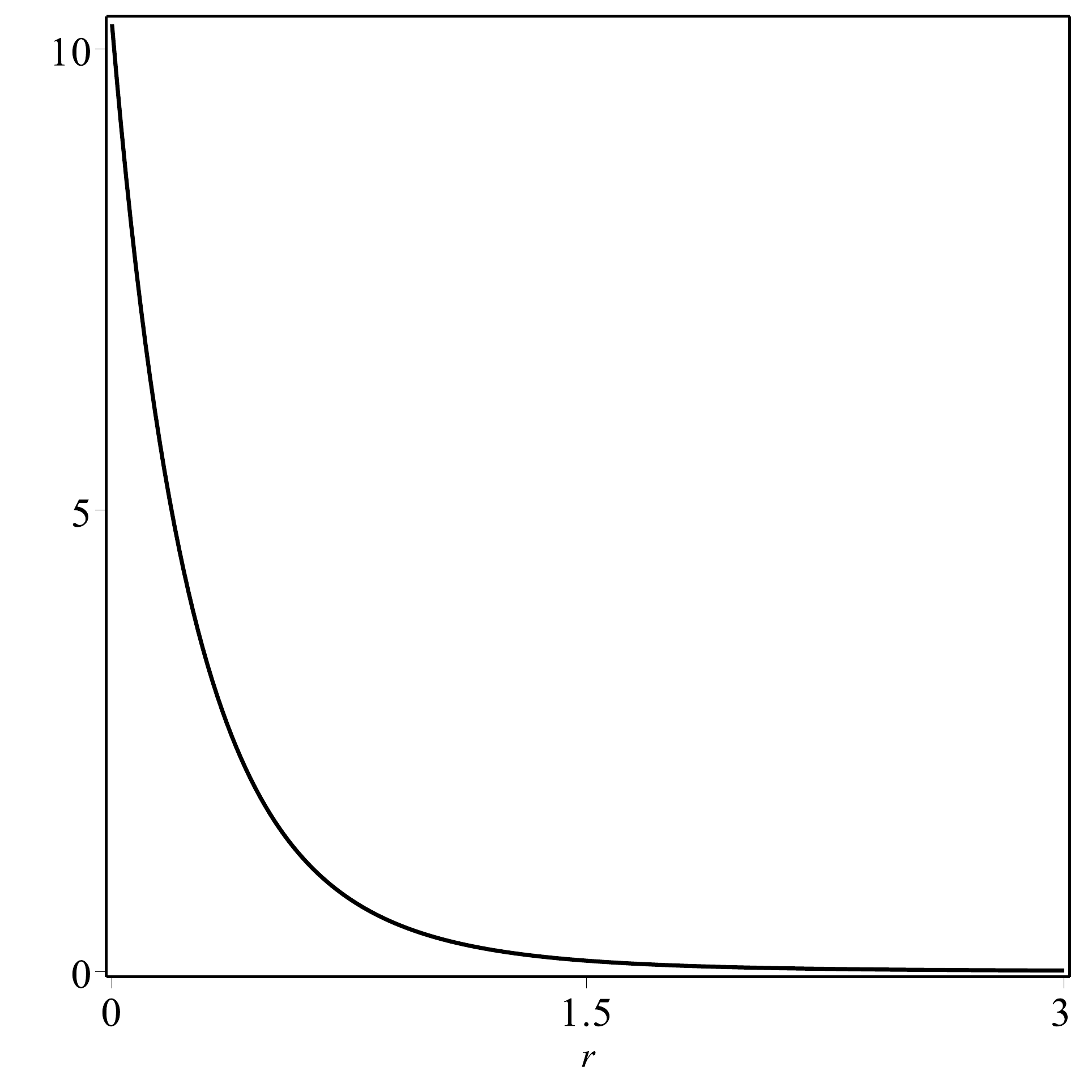}
\caption{In the left panel, we show the solutions $\K(r)$ (descending line) and $\Hc(r)$ (ascending line) that appear in Eq.~\eqref{sols}, and in the right panel we display the energy density \eqref{rhos}.}
\label{fig4}
\end{figure} 
%%%%%%%%%%%%%%%%%%%%%%%

%%%%%%%%%%%%%%%%%%%%%%%
\begin{figure}[t!]
\centering
\includegraphics[width=4cm]{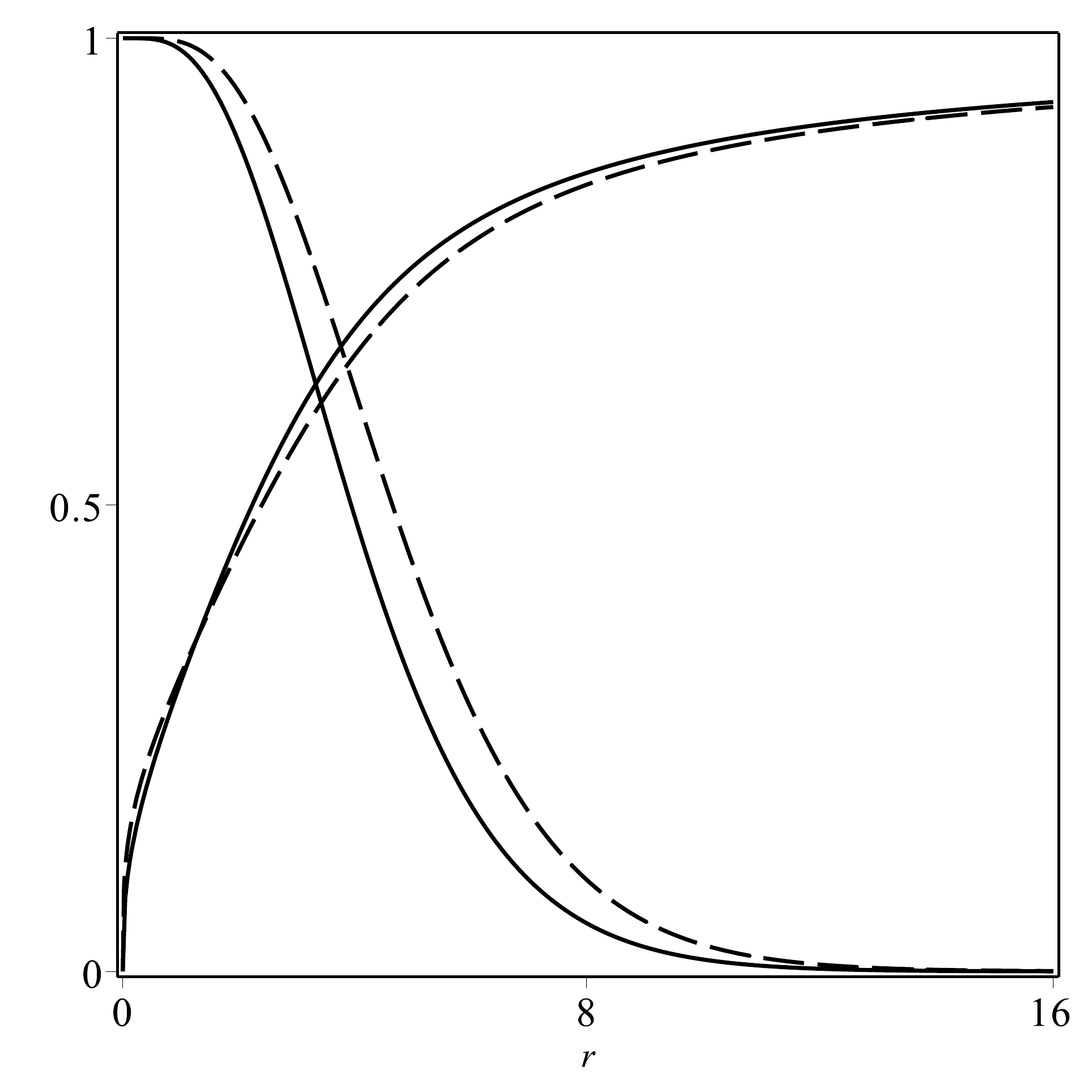}
\includegraphics[width=4cm]{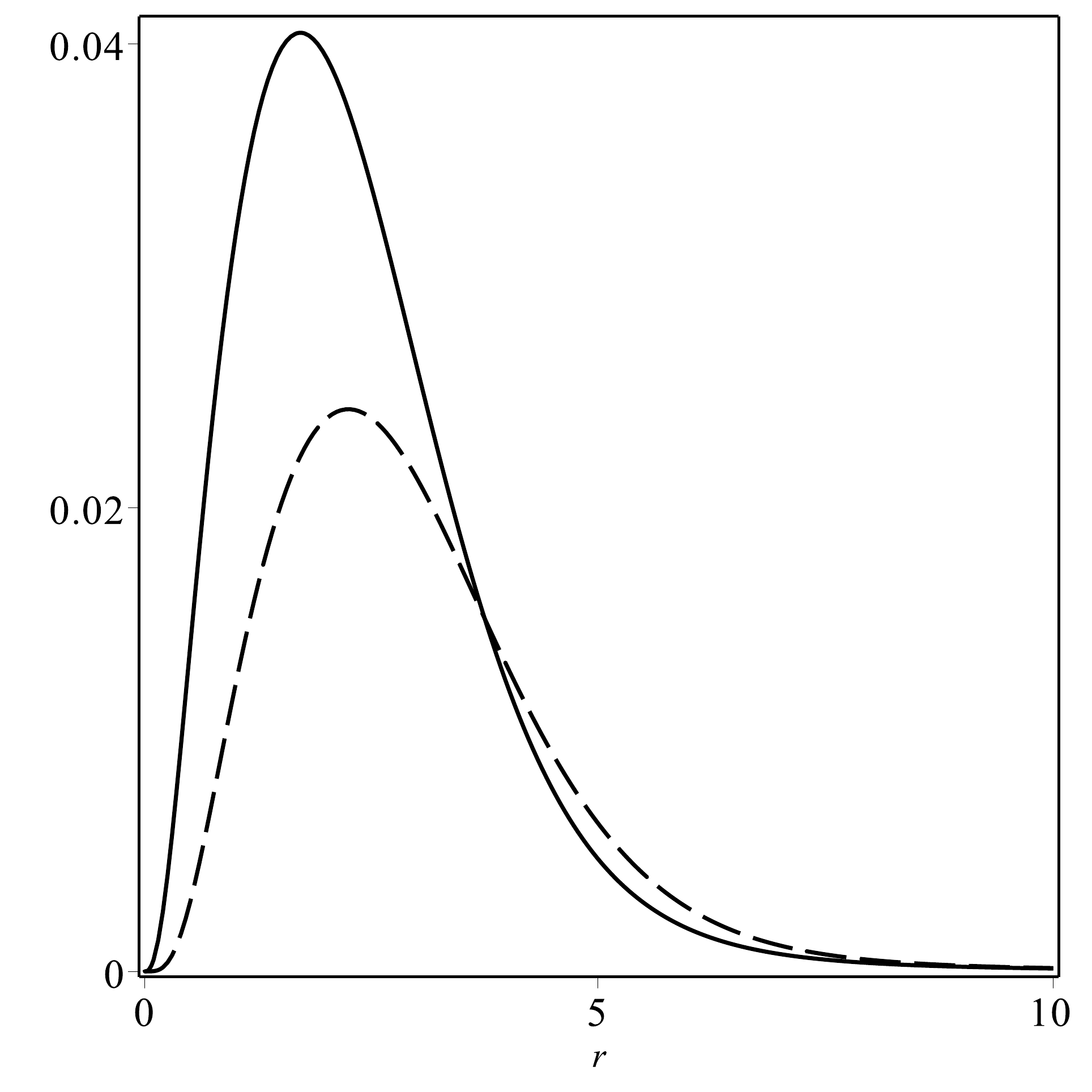}
\caption{In the left panel, we show the solutions $K(r)$ (descending line) and $H(r)$ (ascending line) of Eqs.~\eqref{fov2}, and in the right panel we display the related energy density \eqref{rhoc2}. The solid lines represent the case $\alpha=2$ and the dashed ones stand for $\alpha=3$.}
\label{fig5}
\end{figure} 
%%%%%%%%%%%%%%%%%%%%%%%

\subsection{Shell on small core monopole}
The second example is implemented by considering the core monopole to be the small monopole introduced in Ref.~\cite{Ba3}. We take $\Pc(|\chi|) = |\chi|$ and  $q=w=1$; in this case, the first order Eqs.~\eqref{foh} become
\bes
\bal
\Hc^\prime &= \frac{\Hc(1-\K^2)}{r^2},\\
\K^\prime &=- \K,
\eal
\ees
These equations admit the solutions
\be\label{sols}
\Hc(r) = e^{-\frac{e^{2r}-1}{r\,e^{2r}} - 2\textrm{Ei}(1,2r)} \quad\text{and}\quad \K(r) = e^{-r},
\ee
where $\rm{Ei}(a,z)$ denotes the exponential integral function. The energy density \eqref{rhod} for the core structure is now given by
\be\label{rhos}
\rho_c(r)=\frac{\left(e^{2r}-1\right)^2 + 2r^2e^{2r}}{r^4}\,  e^{-\frac{1 + 2r\,\textrm{Ei}(1,2r) + 4r^2-e^{2r}}{r}}.
\ee
In Fig.~\ref{fig4}, we depict the solutions \eqref{sols} and the above energy density. We see that the energy density presents a peak at the origin and decays very rapidly; this behavior gives rise to the small monopole \cite{Ba3}.

\begin{figure}[t]
\centering
\includegraphics[width=5cm]{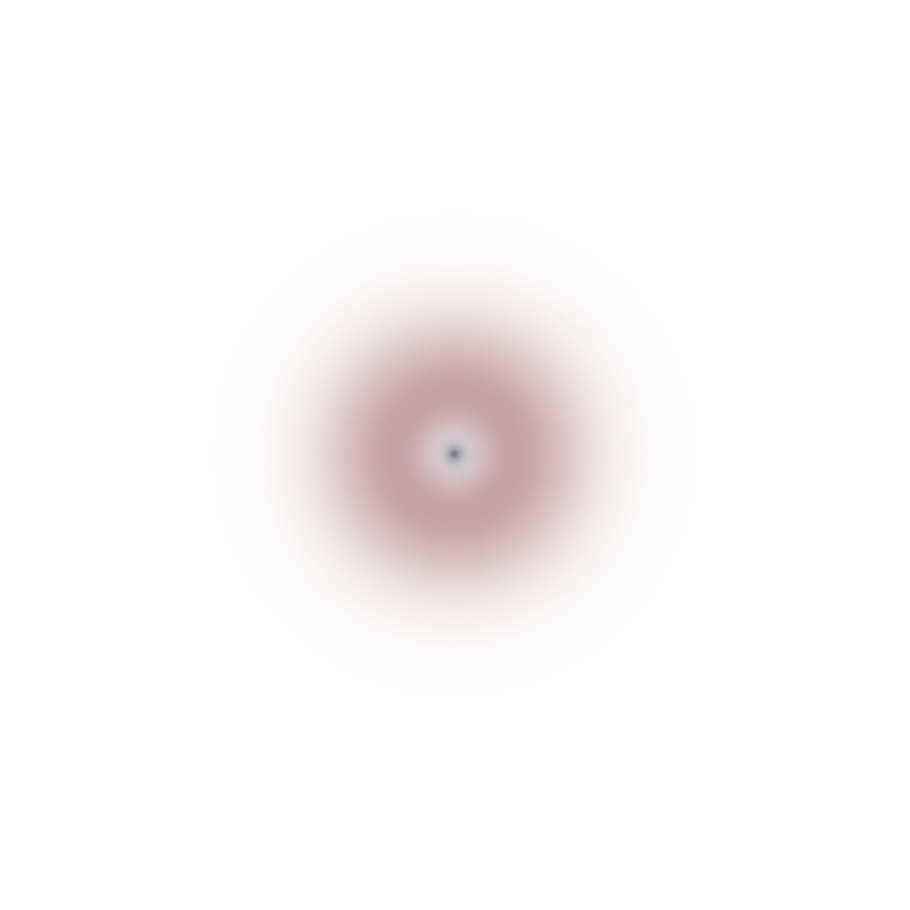}
\caption{A planar section of the energy density passing through the center of the structure. The color blue describes the core of the structure, the small monopole with the energy density shown in Eq.~\eqref{rhos}. The color red describes the shell that represents the hollow monopole with the energy density shown in Eq.~\eqref{rhoc2}, for $\alpha=2$.}
\label{fig6}
\end{figure}

To investigate the effect of the small monopole in the other structure, we take the same magnetic permeability shown in Eq.~\eqref{p1}. In this case, the first order equations \eqref{fov} become
\bes\label{fov2}
\bal
H^\prime &= \frac{(1-K^2)}{r^2 e^{-\frac{\alpha (e^{2r}-1)}{r\,e^{2r}} - 2\alpha\textrm{Ei}(1,2r)}},\\
K^\prime &= - HK e^{-\frac{\alpha (e^{2r}-1)}{r\,e^{2r}} - 2\alpha\textrm{Ei}(1,2r)},
\eal
\ees
As in the previous case, we have been unable to find analytical solutions for the above equations. However, it is possible to estimate their behavior around the origin by considering  $H(r)= H_o(r)$ and $K(r)=1-K_o(r)$. In this case, the leading behavior indicates that  $H_o\propto r^{\left(\sqrt{4\alpha^2 +4\alpha+9} - 2\alpha-1\right)/2}$ and  $K_o\propto r^{\left(\sqrt{4\alpha^2 +4\alpha+9} + 2\alpha+1\right)/2}$. A similar procedure can be done for the asymptotic behavior. By considering $H(r) = 1-H_{asy}(r)$ and $K(r) = K_{asy}$, we get $H_{asy}\propto r^{-1}$ and $K_{asy}(r) \propto r^\alpha e^{-r}$. The energy density can be calculated from Eq.~\eqref{rhoc}; it gives
\be\label{rhoc2}
\begin{split}
\rho_s(r) &= \frac{2{K^\prime}^2}{r^2e^{-\frac{\alpha (e^{2r}-1)}{r\,e^{2r}} - 2\alpha\textrm{Ei}(1,2r)}}\\
&\hspace{4mm}+ {H^\prime}^2e^{-\frac{\alpha (e^{2r}-1)}{r\,e^{2r}} - 2\alpha\textrm{Ei}(1,2r)}\\
&= \frac{2H^2K^2e^{-\frac{\alpha (e^{2r}-1)}{r\,e^{2r}} - 2\alpha\textrm{Ei}(1,2r)}}{r^2} \\
&\hspace{4mm}+ \frac{(1-K^2)^2}{r^4e^{-\frac{\alpha (e^{2r}-1)}{r\,e^{2r}} - 2\alpha\textrm{Ei}(1,2r)}}.
\end{split}
\ee
One can use the behavior of the solutions around the origin to identify the behavior $\rho_s(r\approx0) \approx r^{\sqrt{4\alpha^2 + 4\alpha + 9}-3}$. This shows that the parameter $\alpha$ plays the same role as in the previous example. Moreover, in Fig.~\ref{fig5} we depict the solutions of Eqs.~\eqref{fov2} and the above energy density for $\alpha=2$ and $3$. We note that the energy density presents a hole around the origin, so it also represents a hollow monopole. Since the small monopole is concentrated in a small region, it is nicely nested inside the hollow monopole, as we show in Fig.~\ref{fig6}.
%%%%%%%%%%%%%%%%%%%%%%%
%%%%%%%%%%%%%%%%%%%%%%%

%%%%%%%%%%%%%%%%%%%%%%%
\begin{figure}[t!]
\centering
\includegraphics[width=4cm]{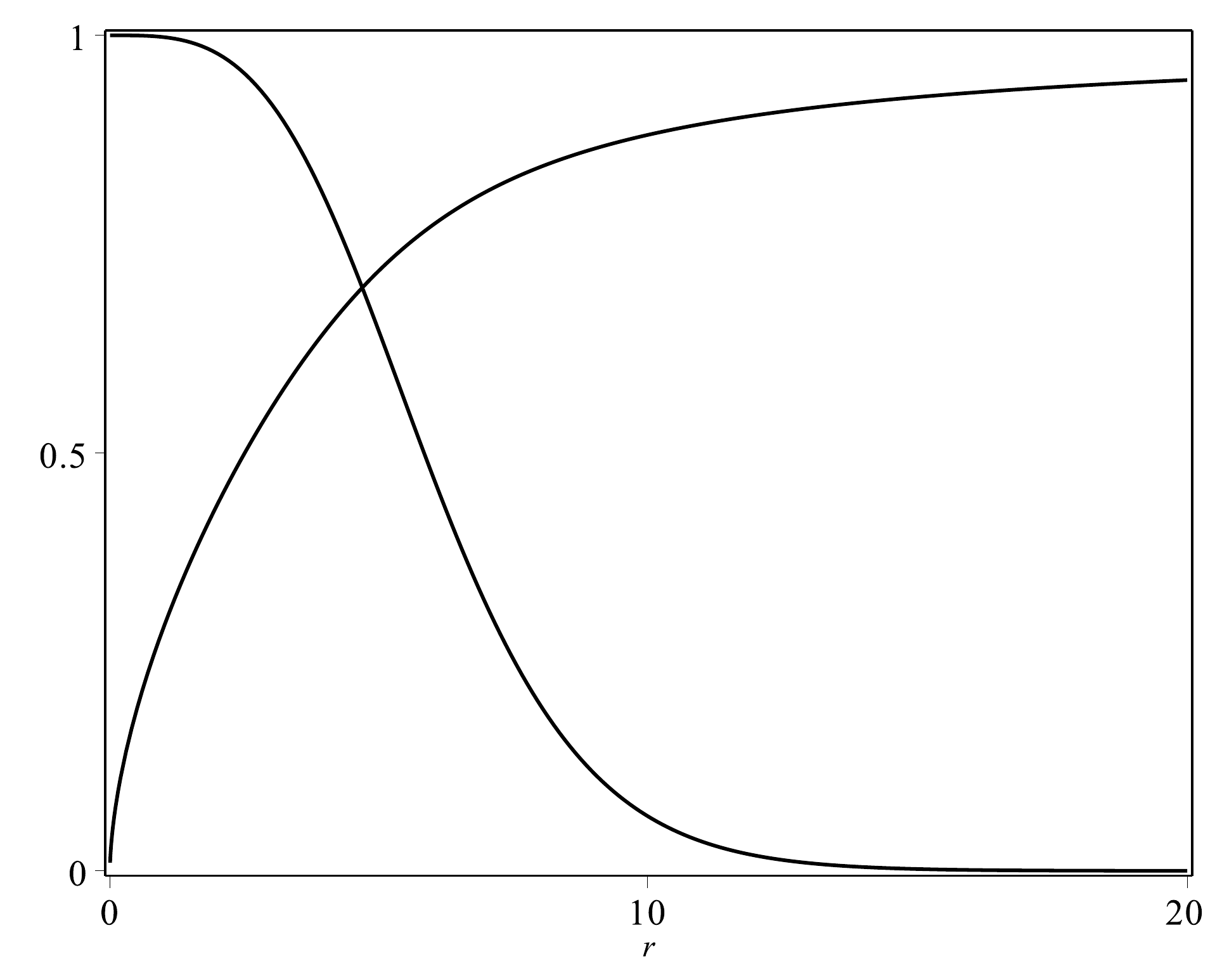}
\includegraphics[width=4cm]{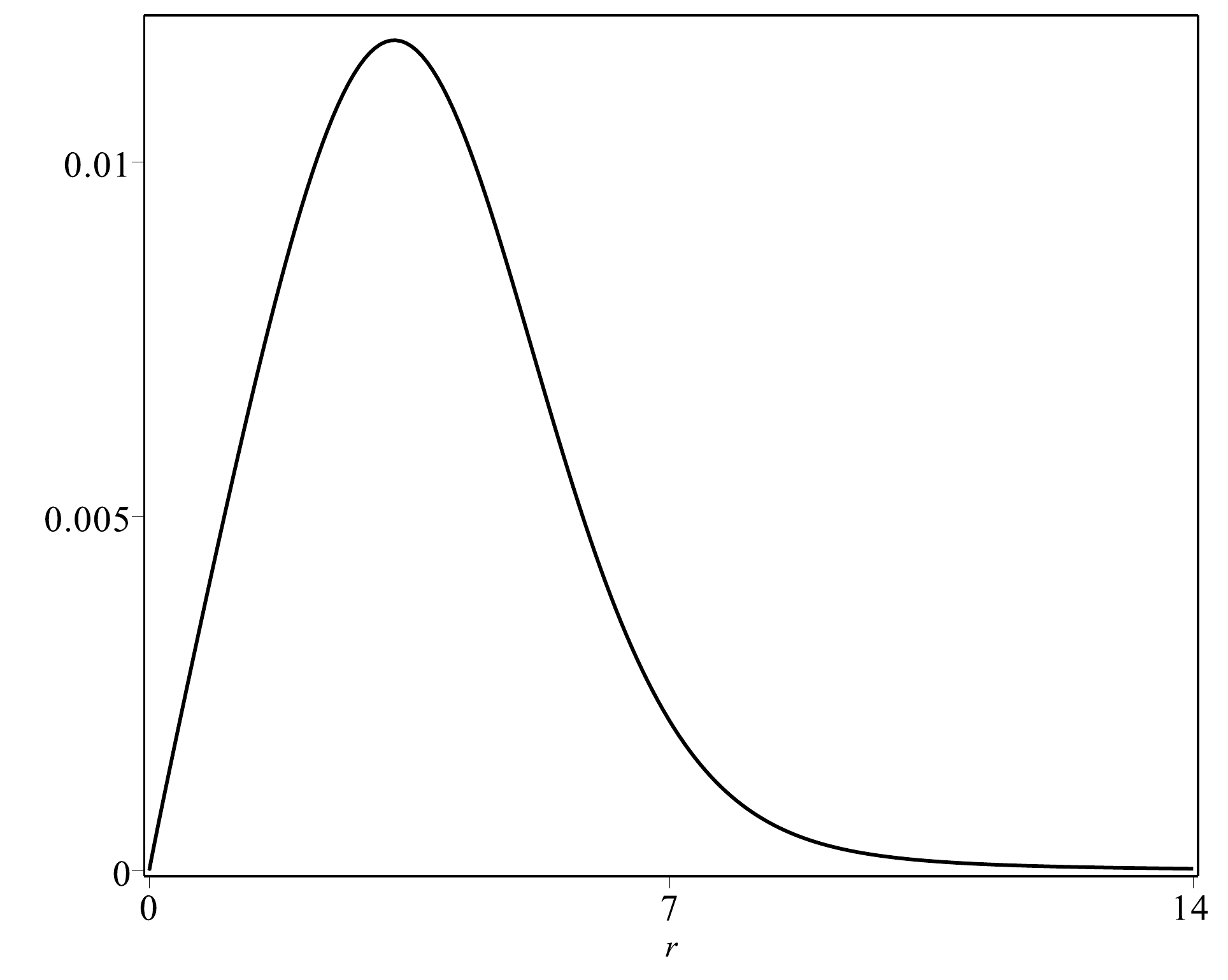}
\caption{In the left panel, we show the solutions $\K(r)$ (descending line) and $\Hc(r)$ (ascending line) of Eq.~\eqref{foshell1}. In the right panel we display the energy density \eqref{rhod} for the magnetic permeability \eqref{pshell1} with $\alpha=3$.}
\label{fig7}
\end{figure} 
%%%%%%%%%%%%%%%%%%%%%%%

\subsection{Shell on shell monopole}

We now turn attention to the construction of a magnetic monopole where a magnetic shell is located on top of another magnetic shell, leaving its core empty. We think that this possibility may be ease to be realized experimentally, since the empty core contributes to circumvent the presence of singularity at the center of the structure that can be constructed with the use of technics similar to that already used to engineer bimagnetic nanoparticles.

We investigate the possibility of constructing a magnetic structure with a shell on top of another shell starting with the choice 
\be\label{pshell1}
\Pc(|\chi|)=1/|\chi|^\alpha.
\ee
This, for $q=w=1$, leads us with the first order equations
\bes\label{foshell1}
\bal
\Hc^\prime &=\frac{(1-\K^2)}{\Hc^\alpha r^2},\\
\K^\prime &=-\Hc^{1+\alpha}\K,
\eal
\ees
Unfortunately, we have not been able to find their explicit solutions. An investigation for small and large values of $r$ is also possible, but we omit it here since it follows as in the previous cases. Instead, we implement a numerical simulation and plot the solutions $\Hc(r)$ and $\K(r)$ in Fig.~\ref{fig7} for $\alpha=3$.
The energy density of the inner shell monopole can be calculated by substituting the numerical solutions in Eq.~\eqref{rhod}; it is also shown in Fig.~\ref{fig7}.

%%%%%%%%%%%%%%%%%%%%%%%
\begin{figure}[t!]
\centering
\includegraphics[width=4cm]{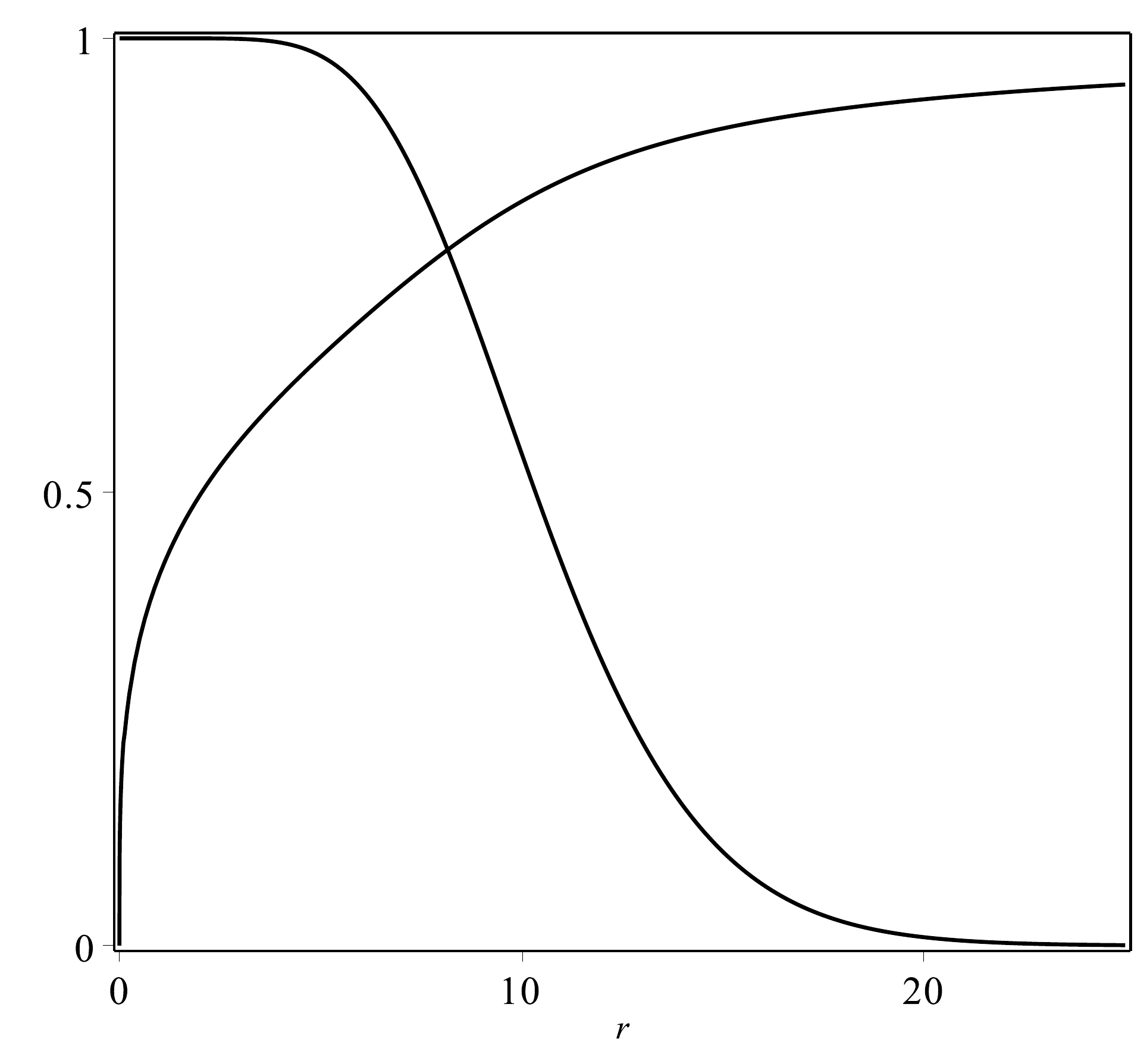}
\includegraphics[width=4cm]{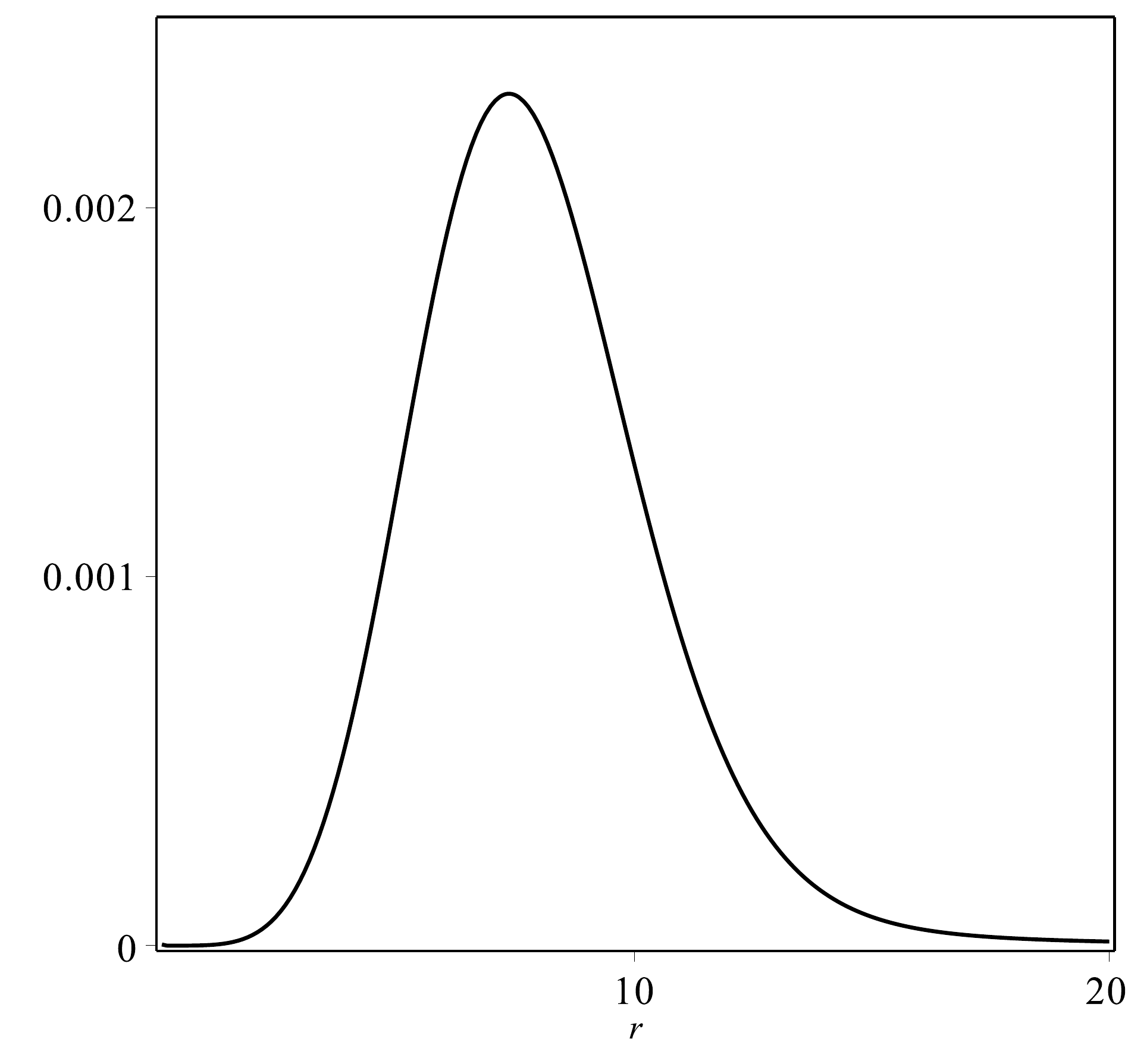}
\caption{In the left panel, we show the solutions $K(r)$ (descending line) and $H(r)$ (ascending line) of Eq.~\eqref{foshell2}. In the right panel we display the energy density \eqref{rhoc} for the magnetic permeability \eqref{pshell2} with $\beta=10$.}
\label{fig8}
\end{figure} 
%%%%%%%%%%%%%%%%%%%%%%%

We then use the solutions of the first order equations \eqref{foshell1} in Eqs.~\eqref{fov}, with the magnetic permeability 
\be\label{pshell2}
P(|\chi|)=1/|\chi|^\beta,
\ee
to get
\bes\label{foshell2}
\bal
H^\prime &=\frac{(1-K^2)}{\Hc^\beta r^2},\\
K^\prime &=-H \Hc^\beta K.
\eal
\ees
Since $\Hc(r)$ is numerical, the solutions of the above equations can only be obtained numerically. The energy density of the outer shell monopole is then obtained from Eq.~\eqref{rhoc}. In Fig.~\ref{fig8}, we plot the solutions and the energy density of the outer shell for $\beta=10$.

\begin{figure}[t]
\centering
\includegraphics[width=5cm]{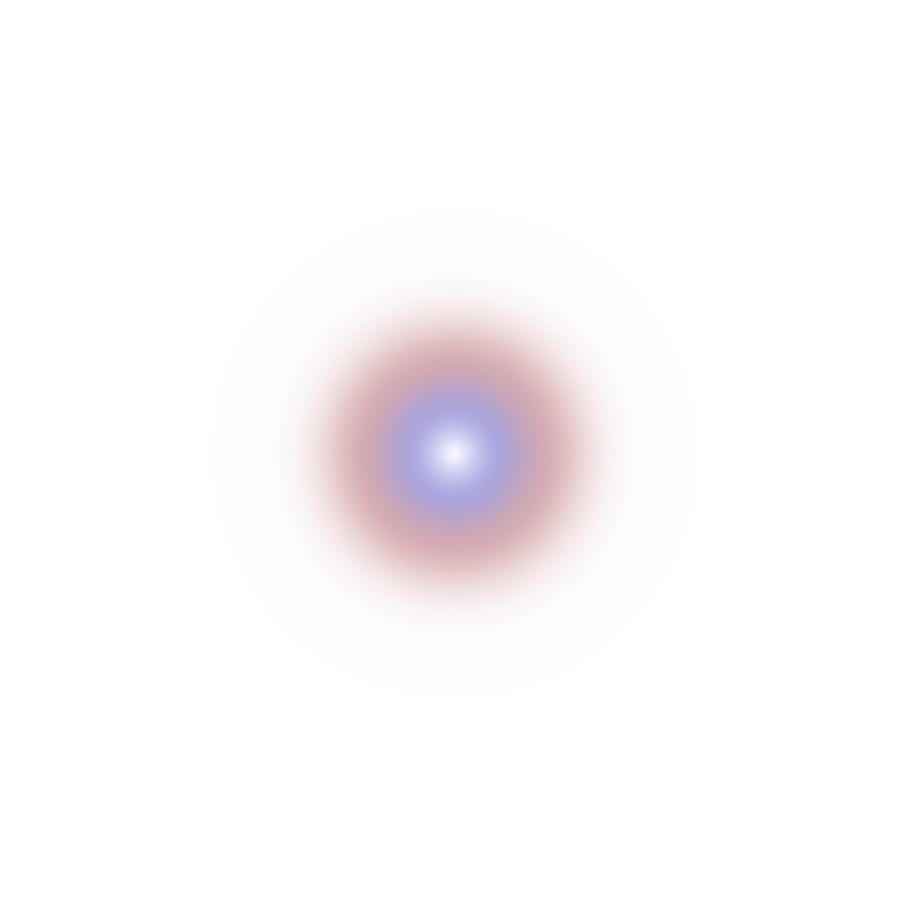}
\caption{A planar section of the energy density passing through the center of the structure. The color blue describes the inner shell with the energy density shown in Eq.~\eqref{rhod} with the magnetic permeability in Eq.~\eqref{pshell1} for $\alpha=3$. The color red describes the outer shell with the energy density shown in Eq.~\eqref{rhoc}, with the magnetic permeability in Eq.~\eqref{pshell2} for $\beta=10$.}
\label{fig9}
\end{figure}

As in the previous cases, we depict a planar section of the structure passing through its center. It is displayed in Fig.~\ref{fig9}, and shows a red shell on top of a blue shell, with an empty core around the origin.

%%%%%%%%%%%%%%%%%%%%%%%
%%%%%%%%%%%%%%%%%%%%%%%

%%%%%%%%%%%%%%%%%%%%%%%%%
\section{Ending comments}
\label{sec:end}

We have described a procedure to obtain bimagnetic structures in a non Abelian model similar to the standard model \cite{thooft,polyakov} used to describe magnetic monopoles in high energy physics. The structures that we found are composed of two distinct monopoles and represent bimagnetic configurations, with the two portions having distinct magnetic behavior. We have examined the cases with a shell on top of a standard monopole, a shell on top of a small monopole and also the case of a shell on top of another shell. The last possibility engenders an empty core, so from the practical point of view it naturally avoids singularity at the origin, and may be advantageous to be prepared using technics of thin films and nanogranular materials available experimentally; see, e.g., Refs.~\cite{Bi1,Bi2,Bi3,bookbi}.

These bimagnetic structures are of current interest and can be used in several contexts, in particular in the case of models beyond the Standard Model, where the left-right symmetry play a role. The left-right symmetry may appear in models of Grand Unified Theories, for example, in the case of the $SO(10)$ model, that can undergo the breaking  $SO(10)\to SU(3)\times SU(2)\times SU(2)\times U(1)$, before becoming the Standard Model. The interest in the left-right symmetry in recent years is due to the fact that experiment at the LHC could confirm it; see, e.g., the recent works \cite{LR11,LR12} and references therein. 

The composed structures can also be used in applications in systems with two distinct condensed phases, and in other circumstances. Some specific possibilities are now being examined, one concerning the addition of fermions to study the existence of fermionic zero modes attached to the bimagnetic structure (see, e.g., \cite{M5} and references therein) and another one, dealing with the behavior of the bimagnetic structure under the action of external fields. Another interesting possibility refers to the extension of the model to the case of three distinct sets of gauge and scalar fields that evolve under the $SU(2)\times SU(2)\times SU(2)$ symmetry. This idea leads to structures that nest three distinct substructures and can ultimately be generalized to the case of multilayer structures, which are also of interest to the characterization of onion-like magnetic nanoparticles with two, three, or four components \cite{Bi3}, and also to study the multi-component nature of color superconductivity.
 
%%%%%%%%%%%%%%%%%%%%%%%%%
\acknowledgements{The authors acknowledge the Brazilian agency CNPq for financial support. DB acknowledges support from Grant No. 306614/2014-6, MAM acknowledges support from Grant No. 140735/2015-1, and RM acknowledges support from Grant No. 306826/2015-1.}
%%%%%%%%%%%%%%%%%%%%%%%%%%%%%%%%%


\begin{thebibliography}{99}
\bibitem{thooft}G. '\! t Hooft, \emph{Nucl. Phys.} B {\bf79}, 276 (1974).
\bibitem{polyakov}A.M. Polyakov, \emph{JETP Lett.} {\bf20}, 194 (1974).
\bibitem{ps} M.K. Prasad and C.M. Sommerfield, \emph{Phys. Rev. Lett.} {\bf35}, 760 (1975).
\bibitem{bogo} E.B. Bogomol'nyi, \emph{Sov. J. Nucl. Phys.} {\bf24}, 449 (1976).
\bibitem{V}A. Vilenkin and E.P.S. Shellard, \emph{Cosmic Strings and
other Topological Defects} (Cambridge University Press, Cambridge, 2000).
\bibitem{MS}N. Manton and P. Sutcliffe, \emph{Topological Solitons} (Cambridge University Press, Cambridge, 2004).
\bibitem{Sh}Y.M. Shnir, \emph{Magnetic Monopoles} (Springer, NewYork, 2005).
\bibitem{W}E. Witten, \emph{Nucl. Phys.} B {\bf249}, 557 (1985).
\bibitem{Shi}M. Shifman, \emph{Phys. Rev.} D {\bf87}, 025025 (2013).
\bibitem{HS}A. Haber and A. Schmitt, \emph{J. Phys.} G 45, 065001 (2018).
\bibitem{M1}M. Shifman, G. Tallarita, and A. Yung, \emph{Phys. Rev.} D {\bf91}, 105026 (2015).
\bibitem{M2}T. Vachaspati, \emph{Phys. Rev. Lett.} {\bf117}, 181601 (2016).
\bibitem{Ba1}D. Bazeia, M.A. Marques and R. Menezes, \emph{Phys. Rev.} D {\bf96}, 025010 (2017).
\bibitem{M3}E. Yakaboylu, A. Deuchert, and M. Lemeshko, \emph{Phys. Rev. Lett.} {\bf119}, 235301 (2017).
\bibitem{M4}Yi Li and F.D.M. Haldane, \emph{Phys. Rev. Lett.} {\bf120,} 067003 (2018).
\bibitem{Ba2}D. Bazeia, M.A. Marques and R. Menezes, \emph{Phys. Rev.} D {\bf97}, 105024 (2018).
\bibitem{M5}C. Cs\'aki, Y. Shirman, J. Terning, and M. Waterbury, \emph{Phys. Rev. Lett.} {\bf120,} 071603 (2018).
\bibitem{Ba3}D. Bazeia, M.A. Marques, and G.J. Olmo, \emph{Phys. Rev.} D {\bf98}, 025017 (2018).
\bibitem{SI1}C. Castelnovo, R. Moessner, and S. L. Sondhi, \emph{Nature} (London) {\bf451}, 42 (2008).
\bibitem{SI2}M.J. Harris, S.T. Bramwell, D.F. McMorrow, T. Zeiske, and K.W. Godfrey, \emph{Phys. Rev. Lett.} {\bf79}, 2554 (1997).
\bibitem{SI3}A.P. Ramirez, A. Hayashi, R.J. Cava, R.B. Siddharthan, and S. Shastry, \emph{Nature} (London) {\bf399}, 333 (1999).
\bibitem{SI4}S.T. Bramwell and M.J.P. Gingras, \emph{Science} {\bf294}, 1495 (2001).
\bibitem{SI5}D.I. Khomskii, \emph{Nat. Commun.} {\bf3}, 904 (2012).
\bibitem{Bi1}M. Estrader et al., \emph{Nat. Commun.} {\bf4}, 2960 (2013).
\bibitem{Bi2}T. Okada et al., \emph{J. Applied Phys.} {\bf114}, 125304 (2013).
\bibitem{PS}J.C. Pati and A. Salam, Phys. Rev. D {\bf10}, 275 (1974); Erratum, Phys. Rev. D {\bf11}, 703 (1975).
\bibitem{LR1}R.N. Mohapatra and J.C. Pati, Phys. Rev. D {\bf11}, 2558 (1975).
\bibitem{LR2}G. Senjanovi\'c and R.N. Mahapatra, Phys. Rev. D {\bf12}, 1502 (1975). 
\bibitem{D1}D. Bazeia, M.J. dos Santos, and R.F. Ribeiro, Phys. Lett. A {\bf208}, 84 (1995).
\bibitem{D2}M.A. Shifman and M.B. Voloshin, Phys. Rev. D {\bf57}, 2590 (1998). 
\bibitem{D3}D. Bazeia and F.A. Brito, Phys. Rev. D {\bf62}, 101701(R) (2000).
\bibitem{Sut}P. Sutcliffe, Phys. Rev. D {\bf68}, 085004 (2003).
\bibitem{Shif}A. Peterson, M. Shifman, G. Tallarita, Ann. Phys. {\bf363}, 515 (2015).
\bibitem{Ba4}D. Bazeia, M.A. Marques, and R. Menezes, Phys. Lett. B {\bf780}, 485 (2018).
\bibitem{Bi3}G. Salazar-Alvarez et al., \emph{J. Am. Chem. Soc.} {\bf133}, 16738 (2011).
\bibitem{bookbi}S.K. Sharma, Editor, \emph{Complex Magnetic Nanostructures} (Springer, 2017).
\bibitem{LR11}P. S. Bhupal Dev, R.N. Mohapatra, and Y. Zhang, Phys. Rev. D {\bf95}, 115001 (2017).
\bibitem{LR12}F.F. Deppisch, T.E. Gonzalo, and L. Graf, Phys. Rev. D {\bf96}, 055003 (2017).
\end{thebibliography}
\end{document}